\documentclass[a4paper,conference]{IEEEtran}
\IEEEoverridecommandlockouts
\usepackage{cite}
\usepackage{amsmath,amssymb,amsfonts}
\usepackage{algorithmic}
\usepackage{graphicx}
\usepackage{textcomp}
\usepackage{xcolor}
\usepackage[colorlinks=true,linkcolor=black,citecolor=black, urlcolor=black]{hyperref}
\hypersetup{
  pdfauthor   = {Schöning, Julius; Kruse, Niklas},
  pdfkeywords = {AI Compliance, AI Act, Trustworthy AI, Explainable AI (XAI), Data Set Compliance
},
  pdftitle    = {Compliance of AI Systems},
  pdfdisplaydoctitle = true
}
\def\BibTeX{{\rm B\kern-.05em{\sc i\kern-.025em b}\kern-.08em
    T\kern-.1667em\lower.7ex\hbox{E}\kern-.125emX}}

\usepackage{balance}
\usepackage{tikz}
\usetikzlibrary{positioning, calc, shapes.geometric, shapes.symbols, arrows.meta, fit, backgrounds, intersections}
\usepackage{pgfplots}
\pgfplotsset{compat=1.16}
\usepackage{pgfplotstable}
\usepackage{array}
\usepackage{multirow}

\usepackage{stfloats}

\usepackage{xstring}

\ifCLASSOPTIONcompsoc
  \usepackage[caption=false,font=normalsize,labelfont=sf,textfont=sf]{subfig}
\else
  \usepackage[caption=false,font=footnotesize]{subfig}
\fi

\usetikzlibrary{external}
\begin{document}

\title{\vspace{2.5cm} Compliance of AI Systems}

\author{\IEEEauthorblockN{Julius Schöning}
\IEEEauthorblockA{\textit{Faculty of Engineering and Computer Science} \\
\textit{Osnabrück University of Applied Sciences}\\
Osnabrück, Germany \\
j.schoening@hs-osnabrueck.de \vspace{-.6cm}
}
\and
\IEEEauthorblockN{Niklas Kruse}
\IEEEauthorblockA{\textit{Faculty of Engineering and Computer Science} \\
\textit{Osnabrück University of Applied Sciences}\\
Osnabrück, Germany \\
niklas.kruse@hs-osnabrueck.de \vspace{-.6cm}}
}

\maketitle
\begin{abstract}
The increasing integration of artificial intelligence (AI) systems in various fields requires solid concepts to ensure compliance with upcoming legislation. This paper systematically examines the compliance of AI systems with relevant legislation, focusing on the EU's AI Act and the compliance of data sets. The analysis highlighted many challenges associated with edge devices, which are increasingly being used to deploy AI applications closer and closer to the data sources. Such devices often face unique issues due to their decentralized nature and limited computing resources for implementing sophisticated compliance mechanisms. By analyzing AI implementations, the paper identifies challenges and proposes the first best practices for legal compliance when developing, deploying, and running AI. The importance of data set compliance is highlighted as a cornerstone for ensuring the trustworthiness, transparency, and explainability of AI systems, which must be aligned with ethical standards set forth in regulatory frameworks such as the AI Act. The insights gained should contribute to the ongoing discourse on the responsible development and deployment of embedded AI systems.
\end{abstract}


\section{Introduction}
In recent years, artificial intelligence (AI) solutions have gained significant attention within scientific discourse across various fields, including agriculture~\cite{Liu2020,Schoning2021}, closed-loop control systems~\cite{Schoening2023,Walczuch2022}, visual analytics~\cite{Wang2024,Tanisaro2015} and automated driving~\cite{Kuznietsov2024,Li2024}. AI opens up entirely new application areas and offers notable application efficiency improvements. Initially conceptual and confined to research test environments, AI-based systems' advancements are increasingly transitioning into economic contexts~\cite{Nortje2020}. This trend is evident in various technologies, where numerous AI-integrated products are commercially available from various manufacturers~\cite{Anderson2022,Parikh2025}.

As a result of this shift, AI solutions are no longer influenced solely by research and academic impact factors but are increasingly subject to societal impact mechanisms~\cite{Cath2017}. Technical progress in AI is not just an influencing factor; subjective elements like the trustworthiness of commercially deployed AI systems have become crucial in their economic utilization~\cite{Ferrario2019}. This importance is underscored by growing mistrust among broad segments of society~\cite{Crockett2020}, which could impede the adoption of new AI technologies.

In response to technological advancements and rising skepticism, European legislators have begun to adapt existing regulations to encompass AI and have introduced new legislation, such as the AI Act~\cite{EuropeanCommission2020}. Even before these societal changes, the field of explainable AI (XAI) was exploring ways to elucidate the actions of AI systems. This pursuit aims to enhance trustworthiness and better understand errors for increased efficiency~\cite{Cyras2021}.

Based on the six steps of the pipeline of applying AI, this paper examines which steps of regulatory requirements influence AI systems and how AI systems need to be extended to become truly trustworthy embedded AI systems. To achieve this, a platform that integrates scientific insights from XAI with regulatory demands as a first best practices approach is introduced.

\section{Steps to Ensure Compliance in applying AI}\label{sec:pipeline}
\begin{figure*}[t!]
  \centering
  \includegraphics[width=0.95\linewidth]{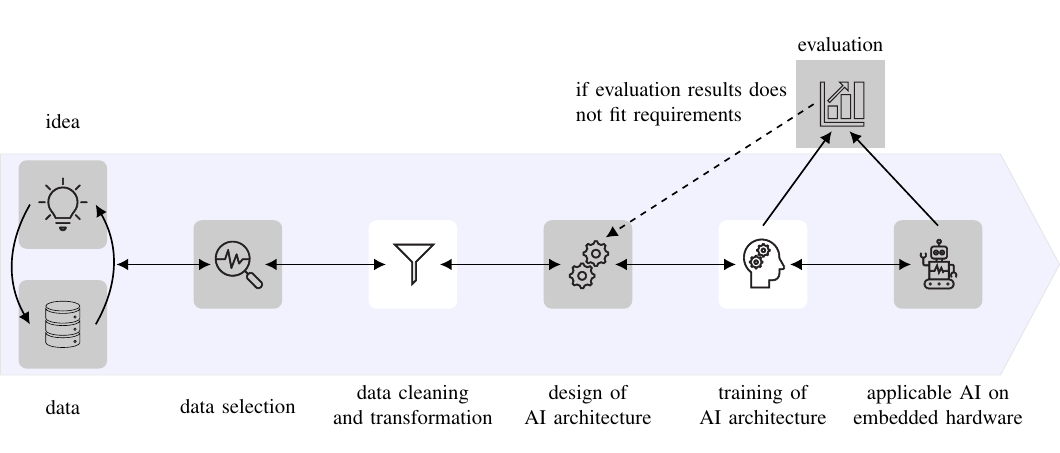}
\caption{The pipeline of applying AI~\cite{Schoening2023a} and the steps where compliance is needed are highlighted with a gray background.}
\label{fig:pipeline}
\end{figure*}

The development of AI follows a six-step process~\cite{Schoening2023a}, as illustrated in Fig.~\ref{fig:pipeline}:
\begin{itemize}
\item \textbf{1.  Data and Application Idea:} AI development starts with data and an application idea. The idea and data often influence each other, with ideas shaping data collection and data inspiring new ideas.
\item \textbf{2.  Data Selection:} Once the application idea and data lake are established, the next step is selecting the data needed to solve the application. It is crucial to ask if a human can perform the task with the selected data for novel tasks, which can enhance early development success. Later, data selection can be more adventurous, especially for embedded hardware.
\item \textbf{3.  Data Cleaning and Transformation:} This step involves cleaning the data to avoid biases and ensure generalization by the AI model. Data transformation is also essential, such as converting time series data into a format suitable for different AI architectures.
\item \textbf{4.  AI Architecture Selection:} This step involves choosing the best applicable AI architecture for the task. With a plethora of options available~\cite{Veen2019}, it is important to carefully examine several architectures. 
\item \textbf{5.  Training the AI Architecture:} The most resource-intensive step involves using 80\% of the data set to train the model. The remaining 10\% is used to evaluate training performance and detect overfitting. 
\item \textbf{6.  Deployment on Embedded Hardware:} The final step is deploying the AI on embedded hardware to solve the desired tasks. Users may test the AI's limits, leading to adversarial attacks~\cite{Huang2020,Fawaz2019}. To mitigate this, the operational design domain should be clearly defined early in the process.
\item \textbf{Evaluation:} The task performance must be evaluated using an unseen Data set, which is usually 10\% of the data set. The AI architecture must be refined and retrained if the performance is unsatisfactory. This cycle is resource-intensive.
\end{itemize}

\section{XAI and Legal Compliance}
\begin{figure*}[b]
\includegraphics[width=\linewidth]{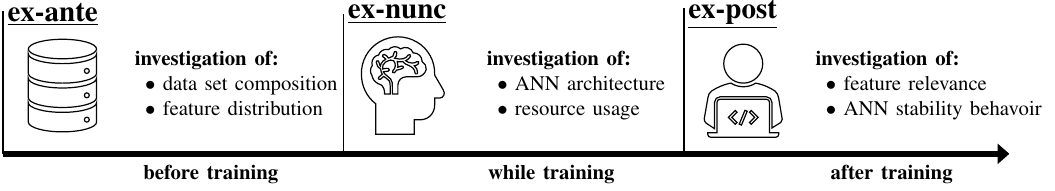}
\caption{Different XAI techniques concerning the training process \cite{Kruse2025}.}
\label{fig:xai}
\end{figure*}

The research topic of XAI has been discussed in recent decades, even before the development of modern AI systems~\cite{Arrieta2020}. However, the advent of powerful AI systems, such as large language models (LLM) and reasoning models, has introduced new approaches and challenges in XAI~\cite{Arrieta2020}. 
XAI methods can be categorized based on the point in the AI model's life cycle at which they are applied~\cite{Cyras2021}, as shown in Fig.~\ref{fig:xai}:

\noindent\textbf{Ex-Ante} These XAI methods are used before training the AI system---step 1 Data and Application Idea. These methods examine the data set to predict the model's behavior based on it. Given the size of modern Data sets, this process is complex but crucial, especially for legal compliance as per Article 10 II of the AI Act, which links the legal conformity of the AI system to the composition of the data set.

\noindent\textbf{Ex-Nunc} These methods focus on the model architectures---step 4---and the training results---step 5---by altering input parameters and observing changes in behavior. They aim to infer the model's decision-making process but risk confusing correlation with causality due to the complexity of modern AI systems.

\noindent\textbf{Ex-Post} These methods examine the model's decision-making, reasoning process, and assumptions made~\cite{Arrieta2020} during operation---step 6 and step evaluation. Theses XAI methods can be useful in confirming and refuting ex-nunc assumptions but are also challenged by the complexity of AI systems.

While XAI does not guarantee trustworthiness on its own, it is a crucial component of trustworthy AI~\cite{Ferrario2019}. Trustworthiness is a subjective concept influenced by explainability, comprehensibility, legality, and imputability~\cite{Hardin2002}. Each of these factors requires the external recipient to understand the AI system's motives, which is essential for assessing legality and ensuring compliance with applicable laws. Therefore, XAI plays a vital role in making AI systems trustworthy, though the specific procedures must be tailored to the target audience and their needs.

\section{Platform-Based Approach for Trustworthy and Compliant AI}
\begin{figure*}[t]
\includegraphics[width=\linewidth]{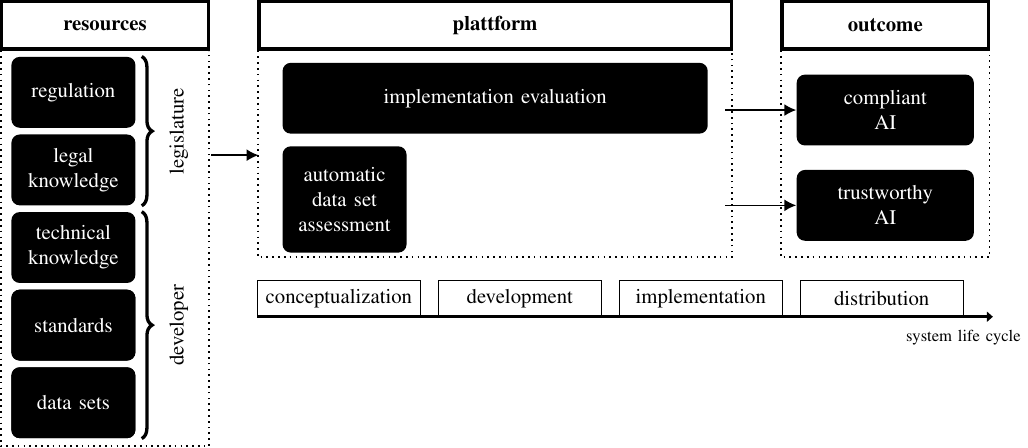}
\caption{Proposed platform structure based on the respective stakeholder resources \cite{Kruse2025}.}
\label{fig:platform}
\end{figure*}

Given the previous considerations, this raises the question of how developments that implement the technological advantages of AI in embedded hardware can be guided to integrate trustworthiness and legal compliance directly within their AI systems. A platform-based approach is proposed as a holistic approach that enables developers of such systems to consider the premises defined here. Within this platform in its proposed structure shown in Fig.~\ref{fig:platform}, the AI-based system and its corresponding stakeholder's needs are sketched by the developers, and the platform will guide and support the developers through the regulations, norms, and standards to ensure the development of a trustworthy and compliant system from the start. Thus, the resulting system will meet the stakeholders' expectations of trustworthiness and comply with the law. 

As the input the technical parameters defined for an AI system are standardized with specific technical requirements as part of such a platform-based concept. The decisive obstacle at this point is that the developers face considerable difficulties in understanding the current legislation, which would have to be adjusted in terms of technical parameters. This lack of understanding of the laws increases the risk that a legal assessment will only occur after conceptualization or even after development. If a legal obstacle is then identified that prevents the solution from being sold on the market, this may mean that the development process for the AI solution has to be partially redone. This late identification of legal obstacles that influence the system architecture delays the development process and increases the costs of AI development. Consequently, it is necessary to carry out a legal assessment at an early stage of development. At the same time, common legal reviews at this development stage are associated with considerable costs. Therefore, the platform solution conceived here should enable the developer to independently carry out an initial legal assessment during the development process. This legal assessment is not intended to replace a professional assessment by an expert but instead to reduce the likelihood of a significant legal problem arising by providing the developer with in-depth knowledge of the legal design of AI systems. For this purpose, an app-based solution that we have previously developed could be used in the platform. This application aims to simulate the process of gathering legal information and then identify potentially critical legal problems. The developer then receives an abstract explanation of this problem and how such problems can be avoided. An expert system is used in the current design and a platform to be developed in the future, rather than an LLM-based solution, which is now more common. One of the reasons for this is that the legal situation in this area of law is changing much more rapidly than in other areas of law. A LLM-based system can face the difficulty that it cannot react quickly enough to selective changes, while an expert system is easier to oversee and adjust. The advantage of an expert system is that it does not fall within the scope of Art. 6 of the AI Act, as it does not act autonomously. If this were the case, then this AI system would qualify as a high-risk AI system, which would entail the risk that the AI system used to promote the trustworthiness of AI is not trustworthy in itself. Moreover, an LLM-based system can only offer limited added value in this application, as the requirements of German profession law have a restrictive effect here. According to the provisions of the Legal Services Act (RDG), legal advice can only be offered within a limited framework and by specific persons. Therefore, the platform presented here should not be used to provide advice in individual cases anyway, but the system's outputs should only be generally applicable. 

The mode of action can be illustrated using an industrial example. The subject of the example used here is an AI system used in a company that assigns workers to an activity using live data from production. Such an AI system presents the developer with several legal obstacles as early as the conception phase, which could potentially trigger a new development. On the one hand, the developer must be aware that this is a high-risk system, as it is an AI system subject to Art. 6 II  and Annex III No. 4 lit. b Tasks distributed to workers. In addition, the AI system also poses a high risk to the fundamental rights of a worker under Art. 2 III, as it can completely deprive a person of work and could thus risk their livelihood. The developer must be aware not only of the fact that this is a high-risk AI system but also of the resulting legal implications. These are manifold, such as the correct risk assessment of the system.

In particular, the requirements of Art. 10 No. 2 lit. f should be mentioned here, according to which the manufacturer must examine the data set to determine whether the composition of the data may influence the health or safety of persons or promote possible prejudices. Concerning the AI system presented here, there could be a risk that the AI system has been trained on incorrect basic data and assigns far too many tasks to an employee at the worksite, thus promoting a harmful effect on the employee's health due to the increased and unreasonable workload. There could also be the problem that the AI system has also been trained with gender-biased data and may, therefore, assign work to women that they are physically unable to perform. The developer can only address the requirements for the data set if he is aware of them before the actual AI system is trained. If the legal assessment only takes place afterward, a new development would have to take place, as the AI system could not be distributed on the market if it were to realize this infringement. Through the proposed expert system as part of the conceptualized platform, it would be possible that the risk of redevelopment would be minimized by the developer being aware of this risk already in the conceptualization phase. Ultimately, the premise would be fulfilled to meet technical parameters correctly with the regulatory requirements without significantly increasing the cost of legal advice or redeveloping the AI system. The structure of the data set has an equally significant influence on the legality of the AI system. If the data set in itself already causes an infringement of a right, like to one's own image, an AI system trained on this data set must be considered untrustworthy. In addition to the complex legal situation, there is the problem of the amount of data to be checked. The proposed platform would, therefore, have to include a method to check the legality of an AI system's database automatically. With regard to image data, such automated data record control has already been implemented in the context of the right to one's own image under Section 22 KUG~\cite{Kruse2024}. 

However, the part of the conceptualized platform presented so far would not yet provide a conclusive statement on the system's trustworthiness. In addition, the platform to be developed must allow the respective recipients to adequately satisfy their needs for trustworthiness. The needs of the different stakeholders are determined by their perspective on the AI system, which ultimately defines which criteria of trustworthiness must be met by the respective proofs. In this respect, the platform to be developed should enable the developers of AI systems, who have previously demonstrated the legality of the AI system, to provide corresponding verification procedures for the various components of trustworthy AI. For example, it should be possible to produce evidence for a supervisory authority, through which the developer can demonstrate in a court hearing that an AI system not only complies with the applicable legal standards as intended but that the legally compliant behavior also satisfies planned motives. Ex-post approaches could be helpful for this, but they should be expanded to include ex-nunc procedures that could be used to substantiate a correlation. These procedures would more vigorously address the concept of legality.

On the other hand, some system users, e.g., need the system to be easy to understand. Proof of trustworthiness should focus more on demonstrating the comprehensibility of the system by showing the user a change in the model's actions using different inputs, if possible, with easy-to-understand visual evidence. Within this step, it is essential to consistently meet the needs of the respective stakeholders to elevate the model from a purely technical-legal concept of trust to an actual level of trust.

Therefore, the proposed platform solution for trustworthy AI should follow a two-part approach. As a first step, it is necessary to combine the technical and legal requirements, for which the developers must already know about the legally correct development of AI systems~\cite{Kruse2024a}. This step can only be successful if the costs of legal advice are reduced, which this platform is not intended to achieve by replacing expert advice altogether but by reducing the likelihood of critical breaches of the law through the AI system. The second approach extends this evidence by further explaining trustworthiness based on the recipients' different conceptual components and expectations.

\section{Conclusion}
As a first best practice solution on compliant AI, this paper introduces a platform-based approach integrating insights from explainable AI (XAI) with regulatory demands. This platform guides developers through the regulatory, normative, and standard requirements, ensuring the development of trustworthy and compliant AI systems from the outset. The platform aims to reduce the risk of legal issues by enabling developers to conduct initial legal assessments during the development process, thereby minimizing the need for costly redevelopments. By addressing the needs of different stakeholders and ensuring that AI systems meet the criteria of trustworthiness, comprehensibility, legality, and imputability, the platform contributes to the responsible development and deployment of AI, fostering broader societal acceptance and adoption of these technologies.

\section*{Acknowledgement}
This paper includes work carried out within the AgrifoodTEF-DE project. AgrifoodTEF-DE (reference: 28DZI04A23) is supported by funds of the Federal Ministry of Agriculture, Food and Regional Identity (BMLEH) based on a decision of the Parliament of the Federal Republic of Germany via the Federal Office for Agriculture and Food (BLE) under the research and innovation program 'Climate Protection in Agriculture'.

\balance
\bibliographystyle{myIEEEtran}
\bibliography{ref.bib}

\begin{thebibliography}{10}
\providecommand{\url}[1]{#1}
\csname url@samestyle\endcsname
\providecommand{\newblock}{\relax}
\providecommand{\bibinfo}[2]{#2}
\providecommand{\BIBentrySTDinterwordspacing}{\spaceskip=0pt\relax}
\providecommand{\BIBentryALTinterwordstretchfactor}{4}
\providecommand{\BIBentryALTinterwordspacing}{\spaceskip=\fontdimen2\font plus
\BIBentryALTinterwordstretchfactor\fontdimen3\font minus
  \fontdimen4\font\relax}
\providecommand{\BIBforeignlanguage}[2]{{%
\expandafter\ifx\csname l@#1\endcsname\relax
\typeout{** WARNING: IEEEtran.bst: No hyphenation pattern has been}%
\typeout{** loaded for the language `#1'. Using the pattern for}%
\typeout{** the default language instead.}%
\else
\language=\csname l@#1\endcsname
\fi
#2}}
\providecommand{\BIBdecl}{\relax}
\BIBdecl

\bibitem{Liu2020}
S.~Y. Liu, ``Artificial intelligence (ai) in agriculture,'' \emph{IT
  Professional}, vol.~22, no.~3, pp. 14--15, May 2020 .
  \doihref{10.1109/mitp.2020.2986121}

\bibitem{Schoning2021}
J.~Schoning and M.~L. Richter, ``Ai-based crop rotation for sustainable
  agriculture worldwide,'' in \emph{2021 IEEE Global Humanitarian Technology
  Conference (GHTC)}.\hskip 1em plus 0.5em minus 0.4em\relax IEEE, Oct. 2021 .
  \doihref{10.1109/ghtc53159.2021.9612460} pp. 142--146.

\bibitem{Schoening2023}
J.~Schöning and H.-J. Pfisterer, ``Safe and trustful ai for closed-loop
  control systems,'' \emph{Electronics}, vol.~12, no.~16, p. 3489, Aug. 2023 .
  \doihref{10.3390/electronics12163489}

\bibitem{Walczuch2022}
D.~Walczuch, T.~Nitzsche, T.~Seidel, and J.~Schoning, ``Overview of closed-loop
  control systems and artificial intelligence utilization in greenhouse
  farming,'' in \emph{2022 IEEE International Conference on Omni-layer
  Intelligent Systems (COINS)}.\hskip 1em plus 0.5em minus 0.4em\relax IEEE,
  Aug. 2022 . \doihref{10.1109/coins54846.2022.9854938} pp. 1--6.

\bibitem{Wang2024}
J.~Wang, X.~Li, C.~Li, D.~Peng, A.~Z. Wang, Y.~Gu, X.~Lai, H.~Zhang, X.~Xu,
  X.~Dong, Z.~Lin, J.~Zhou, X.~Liu, and W.~Chen, ``Ava: An automated and
  ai-driven intelligent visual analytics framework,'' \emph{Visual
  Informatics}, vol.~8, no.~2, pp. 106--114, Jun. 2024 .
  \doihref{10.1016/j.visinf.2024.06.002}

\bibitem{Tanisaro2015}
P.~Tanisaro, J.~Schöning, K.~Kurzhals, G.~Heidemann, and D.~Weiskopf, ``Visual
  analytics for video applications,'' \emph{it - Information Technology},
  vol.~57, no.~1, pp. 30--36, Jan. 2015 . \doihref{10.1515/itit-2014-1072}

\bibitem{Kuznietsov2024}
A.~Kuznietsov, B.~Gyevnar, C.~Wang, S.~Peters, and S.~V. Albrecht,
  ``Explainable ai for safe and trustworthy autonomous driving: A systematic
  review,'' \emph{IEEE Transactions on Intelligent Transportation Systems},
  vol.~25, no.~12, pp. 19\,342--19\,364, Dec. 2024 .
  \doihref{10.1109/tits.2024.3474469}

\bibitem{Li2024}
X.~Li, M.~Zhu, B.~Zhang, X.~Wang, Z.~Liu, and L.~Han, ``A review of artiﬁcial
  intelligence applications in high-speed railway systems,'' \emph{High-speed
  Railway}, vol.~2, no.~1, pp. 11--16, Mar. 2024 .
  \doihref{10.1016/j.hspr.2024.01.002}

\bibitem{Nortje2020}
M.~Nortje and S.~Grobbelaar, ``A framework for the implementation of artificial
  intelligence in business enterprises: A readiness model,'' in \emph{2020 IEEE
  International Conference on Engineering, Technology and Innovation
  (ICE/ITMC)}.\hskip 1em plus 0.5em minus 0.4em\relax IEEE, Jun. 2020 .
  \doihref{10.1109/ice/itmc49519.2020.9198436} pp. 1--10.

\bibitem{Anderson2022}
M.~M. Anderson and K.~Fort, ``From the ground up: developing a practical
  ethical methodology for integrating ai into industry,'' \emph{AI \& SOCIETY},
  vol.~38, no.~2, pp. 631--645, Jul. 2022 .
  \doihref{10.1007/s00146-022-01531-x}

\bibitem{Parikh2025}
N.~A. Parikh, ``Managing ai-first products: Roles, skills, challenges, and
  strategies of ai product managers,'' \emph{IEEE Engineering Management
  Review}, pp. 1--11, 2025 . \doihref{10.1109/emr.2025.3530942}

\bibitem{Cath2017}
C.~Cath, S.~Wachter, B.~Mittelstadt, M.~Taddeo, and L.~Floridi, ``Artificial
  intelligence and the ‘good society’: the us, eu, and uk approach,''
  \emph{Science and Engineering Ethics}, Mar. 2017 .
  \doihref{10.1007/s11948-017-9901-7}

\bibitem{Ferrario2019}
A.~Ferrario, M.~Loi, and E.~Viganò, ``In ai we trust incrementally: a
  multi-layer model of trust to analyze human-artificial intelligence
  interactions,'' \emph{Philosophy \&; Technology}, vol.~33, no.~3, pp.
  523--539, Oct. 2019 . \doihref{10.1007/s13347-019-00378-3}

\bibitem{Crockett2020}
K.~Crockett, M.~Garratt, A.~Latham, E.~Colyer, and S.~Goltz, ``Risk and trust
  perceptions of the public of artifical intelligence applications,'' in
  \emph{2020 International Joint Conference on Neural Networks (IJCNN)}.\hskip
  1em plus 0.5em minus 0.4em\relax IEEE, Jul. 2020 .
  \doihref{10.1109/ijcnn48605.2020.9207654} pp. 1--8.

\bibitem{EuropeanCommission2020}
\BIBentryALTinterwordspacing
{European Commission}. (2020) White paper on artificial intelligence: A
  european approach to excellence and trust. [Online]. Available:
  \url{https://commission.europa.eu/publications/white-paper-artificial-intelligence-european-approach-excellence-and-trust_en}
\BIBentrySTDinterwordspacing

\bibitem{Cyras2021}
K.~Čyras, A.~Rago, E.~Albini, P.~Baroni, and F.~Toni, ``Argumentative xai: A
  survey,'' 2021.

\bibitem{Schoening2023a}
J.~Schöning and C.~Westerkamp, ``Ai-in-the-loop -- the impact of hmi in
  ai-based application,'' 2023.

\bibitem{Veen2019}
\BIBentryALTinterwordspacing
F.~van Veen and S.~Leijnen. (2019, Apr.) The neural network zoo. [Online].
  Available: \url{https://www.asimovinstitute.org/neural-network-zoo/}
\BIBentrySTDinterwordspacing

\bibitem{Huang2020}
X.~Huang, D.~Kroening, W.~Ruan, J.~Sharp, Y.~Sun, E.~Thamo, M.~Wu, and X.~Yi,
  ``A survey of safety and trustworthiness of deep neural networks:
  Verification, testing, adversarial attack and defence, and
  interpretability,'' \emph{Computer Science Review}, vol.~37, p. 100270, 2020
  . \doihref{10.1016/j.cosrev.2020.100270}

\bibitem{Fawaz2019}
H.~I. Fawaz, G.~Forestier, J.~Weber, L.~Idoumghar, and P.-A. Muller,
  ``Adversarial attacks on deep neural networks for time series
  classification,'' in \emph{International Joint Conference on Neural Networks
  ({IJCNN})}.\hskip 1em plus 0.5em minus 0.4em\relax {IEEE}, 2019 .
  \doihref{10.1109/ijcnn.2019.8851936}

\bibitem{Kruse2025}
N.~Kruse and J.~Schöning, ``Explainable and trustworthy ai compliance for
  farms,'' in \emph{45. Jahrestagung der Gesellschaft für Informatik in der
  Land-, Forst- und Ernährungswirtschaft (GIL)}, 2025 .
  \doihref{10.18420/giljt2025\_32}

\bibitem{Arrieta2020}
A.~B. Arrieta, N.~Díaz-Rodríguez, J.~D. Ser, A.~Bennetot, S.~Tabik,
  A.~Barbado, S.~Garcia, S.~Gil-Lopez, D.~Molina, R.~Benjamins, R.~Chatila, and
  F.~Herrera, ``Explainable artificial intelligence (xai): Concepts,
  taxonomies, opportunities and challenges toward responsible ai,''
  \emph{Information Fusion}, vol.~58, pp. 82--115, Jun. 2020 .
  \doihref{10.1016/j.inffus.2019.12.012}

\bibitem{Hardin2002}
R.~Hardin, \emph{Trust and trustworthiness}, ser. The Russell Sage Foundation
  series on trust.\hskip 1em plus 0.5em minus 0.4em\relax New York: Russell
  Sage Foundation, 2002, no. volume 4. ISBN 9781610442718 Includes
  bibliographical references and index.

\bibitem{Kruse2024}
N.~Kruse and J.~Schöning, ``Legal conform data sets for yard tractors and
  robots,'' \emph{Computers and Electronics in Agriculture}, vol. 223, p.
  109106, Aug. 2024 . \doihref{10.1016/j.compag.2024.109106}

\bibitem{Kruse2024a}
N.~Kruse, P.~Wachter, and J.~Schöning, ``Compliance of agricultural ai systems
  – app-based legal verification throughout the development,'' in \emph{44.
  Jahrestagung der Gesellschaft für Informatik in der Land-, Forst- und
  Ernährungswirtschaft (GIL)}, 02 2024 . \doihref{10.18420/giljt2024\_01}

\end{thebibliography}

\end{document}